\documentclass[12pt,aps,floats,showpacs,amssymb,tightenlines]{revtex4}

\usepackage{color}
\usepackage{graphicx}
\usepackage{amsmath}
\usepackage{amsfonts}
\usepackage{amscd}
\usepackage{epsfig}
\usepackage{amssymb}
\usepackage{tabularx}

\begin{document}

\title{The effect of radiative gravitational modes on the dynamics of a cylindrical shell of counter rotating particles}
\author{Reinaldo J. Gleiser} \email{gleiser@fis.uncor.edu} \author{Marcos A. Ramirez}
\affiliation{Instituto de F\'{\i}sica Enrique Gaviola, FAMAF, Universidad Nacional de C\'ordoba,
Ciudad Universitaria, (5000) C\'ordoba, Argentina}
\email{gleiser@fis.uncor.edu}

\begin{abstract}
In this paper we consider some aspects of the relativistic dynamics
of a cylindrical shell of counter rotating particles. In some sense
these are the simplest systems with a physically acceptable matter
content that display in a well defined sense an interaction with the
radiative modes of the gravitational field. These systems have been
analyzed previously, but in most cases resorting to approximations,
or considering a particular form for the initial value data. Here we
show that there exists a family of solutions where the space time
inside the shell is flat and the equation of motion of the shell
decouples completely from the gravitational modes. The motion of the
shell is governed by an equation of the same form as that of a
particle in a time independent one dimensional potential. We find
that under appropriate initial conditions one can have collapsing,
bounded periodic, and unbounded motions. We analyze and solve also
the linearized equations that describe the dynamics of the system
near a stable static solutions, keeping a regular interior. The
surprising result here is that the motion of the shell is completely
determined by the configuration of the radiative modes of the
gravitational field. In particular, there are oscillating solutions
for any chosen period, in contrast with the ``approximately
Newtonian plus small radiative corrections" motion expectation. We
comment on the physical meaning of these results and provide some
explicit examples. We also discuss the relation of our results to
the initial value problem for the linearized dynamics of the shell.
\end{abstract}

\maketitle

\section{Introduction}

In this paper we consider some aspects of the relativistic dynamics
of a cylindrical shell of counter rotating particles. In some sense
these are the simplest systems with a physically acceptable matter
content that display in a well defined sense an interaction with the
radiative modes of the gravitational field. The dynamics of these
systems was analyzed originally by Apostolatos and Thorne
\cite{apostol}, but the evolution was considered in detail only over
very short periods of time, and imposing a particular form for the
initial data, the ``momentarily static radiation free" (MSRF) form
\cite{note1}, and the question of the general evolution in time of
the system has remained largely unexplored. We notice that most of
the literature that followed the work Apostolatos and Thorne has
concentrated in the problem of collapse (see, for instance,
\cite{goncalves} and \cite{wang}), and in general imposing
particular forms for the fields, that may include also some form of
non gravitational radiation outside the shell (see, for instance,
\cite{pereira}, \cite{gleiser} or \cite{seriru}). In a recent paper
Hamity, Barraco and C\'ecere \cite{hamity}, have considered again
the relativistic dynamics of these systems. In particular, since the
system may have stable static configuration, and in the Newtonian
limit small departures form the static configuration lead to
periodic motions, it was expected that in the fully relativistic
dynamics the inclusion of gravitational radiation modes should lead
to a damping of these oscillations, through some form of ``radiation
reaction". This expectation appears to be satisfied in the numerical
solutions obtained in \cite{hamity}. A closer analysis reveals,
however, that the authors assumed an approximation where the back
reaction of the radiative modes is essentially disregarded. This
approximation would be justified if the coupling to the
gravitational radiation modes had only a small effect on the
dynamics of the shell. It turns out, however, as is shown in the
present paper, that rather the opposite situation holds, and the
dynamics is completely dominated by the behaviour of these modes. In
fact we find that, in some sense, the coupling of the shell to the
gravitational radiation modes is as strong as it can be, a
remarkable fact that shows the dynamics of this system cannot be
approximated by a Newtonian dynamics plus post - Newtonian
corrections, as in the case of models where matter is confined to a
bounded region.

The plan of the paper is as follows. After setting up the problem in
Section II, we show in Section III that there exists a family of
solutions where the space time inside the shell is flat and the
equation of motion of the shell decouples completely from the
gravitational modes. The motion of the shell is governed by an
equation of the same form as that of a particle in a time
independent one dimensional potential. We find that under
appropriate initial conditions one can have collapsing, bounded
periodic, or unbounded motions. Next, in Section V we analyze the
linearized equations that describe the dynamics of the system near a
stable static solutions, keeping a regular interior. The surprising
result here is that the motion of the shell is completely determined
by the configuration of the radiative modes of the gravitational
field. In particular, there are oscillating solutions for any chosen
period, in contrast with the ``approximately Newtonian plus small
radiative corrections" motion expectation. Another interesting modes
that appear here are the ``anti resonances" discussed in Section VI.
In Section VII we consider the general behaviuor of the periodic
solutions, and in Section VIII their relation to the initial value
problem for the linearized dynamics of the shell. We comment on the
physical meaning of these results and provide some explicit
examples. We also consider the role of the momentarily static and
radiation free initial data of \cite{apostol}, in this context. Some
closing comments are contained in Section IX.

\section{Equations of motion}

We consider a spacetime $M = M^-\cup \Sigma \cup M^+$ ($M^-$ and$
M^+$ are manifolds with boundary where the boundaries are identified
with the 3-manifold $\Sigma$) with cylindrical symmetry where
$\Sigma$ is the history of a hollow cylinder composed of
counter-rotating particles of rest mass equal to unity; $M^- (M^+)$
is the vacuum interior (exterior) region of the cylinder. In the
vacuum interior $(M^-)$ and exterior $(M^+)$ of the shell, we
introduce canonical cylindrical coordinates $(t, r, z, \phi)$. The
metric takes the form \cite{apostol}.
\begin{equation}\label{eq1}
ds^2_{\pm} = e^{2\gamma_{\pm}-2 \psi_{\pm}}
\left(dr^2-dt^2_{\pm}\right)+ e^{2 \psi_{\pm}}dz^2 +e^{-2
\psi_{\pm}}r^2 d\phi^2
\end{equation}

Dropping the $\pm$ indices, the Einstein field equations in the
empty space inside and outside the shell are,
\begin{equation}\label{eq2}
    \psi_{,rr} +\frac{1}{r}\psi_{,r} -\psi_{,tt} = 0
\end{equation}
\begin{equation}\label{eq3}
\gamma_{,t} = 2 r \psi_{,r} \psi_{,t}\;\;\; , \;\;\; \gamma_{,r} = r
\left[(\psi_{,r})^2+(\psi_{,t})^2\right]
\end{equation}

We may interpret $\psi(r, t)$ as playing the role of a gravitational
field whose static part is the analogue of the Newtonian potential.
The time dependent solutions of (\ref{eq2}) represent gravitational
waves \cite{ER}. Equation (\ref{eq2}) is the integrability condition of
Eqs. (\ref{eq3}). The coordinates $(z, \phi, r )$ and the metric
function $\psi$ are continuous across the shell $\Sigma$, while $t$
and the metric function $\gamma$ are discontinuous. Smoothness of
the spacetime geometry on the axis $r = 0$ requires that $ \gamma =
0$ and $\psi$ finite at $r = 0$. The junction conditions of $M^-$
and $M^+$ through $\Sigma$ require the continuity of the metric and
specify the jump of the extrinsic curvature $K^{\pm}$ compatible
with the stress energy tensor on the shell. The induced metric on
$\Sigma$ is given by
 \begin{equation}\label{eq4}
ds^2_{\Sigma} =-d \tau^2 + e^{2 \psi_{\Sigma}} dz^2 +  e^{-2
\psi_{\Sigma}} R^2 d\phi^2
 \end{equation}

Here $\psi_{\Sigma}(\tau) =
\psi_+(R(\tau),t_+(\tau))=\psi_-(R(\tau),t_-(\tau))$. The evolution
of the shell is characterized by $R(\tau)$, which is the radial
coordinate $r$ at the shell's location and $\tau$ the proper time of
an observer at rest on $\Sigma$. If we assume, as in \cite{apostol},
and \cite{hamity}, that the shells is made up of equal mass counter
rotating particles, the Einstein field equations on the shell may be
put in the form,
\begin{equation}\label{eq5}
\psi^+_{,n}-\psi^-_{,n} = -\frac{2 \lambda}
{  \sqrt{R^2+ e^{2\psi_{\Sigma}} J^2}  }
\end{equation}
\begin{equation}\label{eq6}
X^+-X^- = -\frac{4 \lambda \sqrt{R^2+e^{2 \psi_{\Sigma}}J^2}}{R}
\end{equation}
where the constants $\lambda$ and $J$ are, respectively, the proper
mass per unit Killing length of the cylinder and the angular
momentum per unit mass of the particles. The other quantities in
(\ref{eq5},\ref{eq6}) are given by,
\begin{equation}\label{eq7}
X^{\pm} \equiv \frac{\partial t_{\pm}}{\partial \tau} =
+\sqrt{e^{-2(\gamma_{\pm}-\psi_{\Sigma})} +\dot{R}^2}
\end{equation}
\begin{equation}\label{eq8}
\psi^{\pm}_{,n} = \psi^{\pm}_{,r} X^{\pm} +\psi^{\pm}_{,t} \dot{R}
\end{equation}
where a dot indicates a $\tau$ derivative, and we also have,
\begin{eqnarray}
\label{eq9}
\frac{d^2 R}{d\tau^2} &=&  \dot{R}\dot{\psi_{\Sigma}} - R
\left[(\dot{\psi_{\Sigma}})^2 +(\psi^-_{,n})^2\right]  \nonumber \\
& & +\frac{ R^2 \psi^-_{,n} X^-}{R^2+e^{2 \psi_{\Sigma}}J^2}
-\frac{\lambda R^2 X^-}{(R^2+e^{2 \psi_{\Sigma}}J^2)^{3/2}}
+\frac{J^2 e^{2 \psi_{\Sigma}}X^-X^+}{R(R^2+e^{2 \psi_{\Sigma}}J^2)}
\end{eqnarray}

Equations (\ref{eq5},\ref{eq6},\ref{eq9}), together with
(\ref{eq2},\ref{eq3}) determine the evolution of the shell and of
the gravitational field to which it is coupled. The relevant
functions: $R(\tau)$, $\psi_{\pm}(r,t_{\pm})$, and
$\gamma_{\pm}(r,t_{\pm})$ appear satisfying a rather complex set of
coupled ordinary and partial differential equations, with the
boundary values for $\psi_{\pm}$ and $\gamma_{\pm}$ at $t =
t_{\pm}(\tau),r = R(\tau)$ directly coupled to the motion of the
shell. Because of this complexity, the system was first analyzed in
\cite{apostol} only to show some properties of the motion, although
no solution was obtained, and later in \cite{hamity}, where, after
introducing a second shell, mainly for technical reasons, an
approximation that leads to an effective decoupling of (\ref{eq9})
was used, to avoid considering the complex boundary problem that
results for the wave equations (\ref{eq2}) for $\psi_{\pm}$. Some
full solutions of the problem are considered in the following Sections.

\section{A restricted set of solutions}

A full solution of the problem should provide the evolution of
arbitrary initial data, satisfying the constraints imposed by the
field equations. This is, clearly, a very complex problem. There is,
however, a restriction on the set of solutions that while retaining
its most interesting feature, namely, the coupling of the shell with
radiative modes of the gravitational field, still simplifies
considerably the system, allowing for a complete analysis of the
resulting evolution and of its physical meaning.

\subsection{Static solutions}

We will first consider the static solutions for a shell of constant
radius $R$, assuming an empty flat interior \cite{apostol},
\cite{hamity}. In this case we may take $\gamma_-=0$, $\psi_- =0$,
implying $\psi_{\Sigma}=0$, and for the exterior field we have,
\begin{equation}\label{eq1a}
\psi^+(r)= - \kappa \ln(r/R) \;\;\; , \;\;\; \gamma^+(r)= \gamma_0
+\kappa^2 \ln(r/R)
\end{equation}
then,
\begin{equation}\label{eq2a}
X^+(r)= e^{-\gamma_0} \;\;\; , \;\;\; X^-=1
\end{equation}
Since $\dot{R}=0$, $\dot{\psi}_{\Sigma}=0$, and $\psi^-_{,n}=0$, we
find,
\begin{equation}\label{eq3a}
\kappa = 2 \frac{J^2}{R^2} \;\;\;,\;\;\; \gamma_0 =2
\ln\left[(R^2+2J^2)/R^2 \right]
\end{equation}
and,
\begin{equation}\label{eq4a}
\lambda = \frac{ J^2 R \sqrt{J^2+R^2}}{(2 J^2+R^2)^2}
\end{equation}
 This means that for any static solution we
must have $\lambda \leq 0.15879...$, (see \cite{hamity}).

\subsection{Non static solutions with a flat interior}

The previous results indicate that, at least for the static case, we
have solutions where the interior region of the shell is empty and
flat. We notice that for a similar problem, namely a shell of
counter rotating particles, but with spherical symmetry, we may have
non static solutions where the radius of the shell changes in time,
but the interior remains flat. In this case the spherical symmetry
is crucial, as this implies that there are no radiative modes for
the gravitational field. This is not the case for cylindrical
symmetry, and, in general, one does not expect that in the non
static case the interior will remain flat, because radiative
gravitational modes, corresponding to a non static $\psi$, will in
general penetrate the interior region for, otherwise, the matching
conditions would not be satisfied. Nevertheless, given the existence
of the static solution with an empty flat interior, it is worthwhile
to explore to what extent, if any, this condition can be generalized
to a non static solution. We, therefore, assume again $\gamma_-=0$,
$\psi_- =0$, (implying $\psi_{\Sigma}=0$, $\psi^-_{,n}=0$, and
$X^-=\sqrt{1+(\dot{R})^2}$ ), but place no restriction on either
$R(\tau)$ or $\psi_+$, and $\gamma_+$. The field equations are now
(\ref{eq2}), (\ref{eq3}), and on the shell we have,
\begin{equation}\label{eq5a}
 X^+ =\sqrt{1+(\dot{R})^2}-\frac{2 \lambda}{\sqrt{R^2+J^2}}
\end{equation}
Using this, and the fact that $\dot{\psi}_{\Sigma}=0$, we find,
\begin{equation}\label{eq6a}
\frac{d^2R}{d\tau^2} = \frac{(1+(\dot{R})^2) J^2}{R(R^2+J^2)} -
\frac{\sqrt{1+(\dot{R})^2} (R^2+2J^2)^2 \lambda}{R^2
(R^2+J^2)^{3/2}}
\end{equation}

The first surprising thing about this equation is that it contains
no information on $\psi$, and, therefore, it is an autonomous
equation, completely decoupled from the gravitational mode. Equally
unexpected is that it admits a simple first integral, given by,
\begin{equation}\label{eq7aC}
 C=-\frac{\lambda}{2}\ln  \left( {R}^{2}+{J}^{2} \right) +2\,{\frac {
\lambda\,{J}^{2}}{{R}^{2}}}-{\frac { \sqrt {{R}
^{2}+{J}^{2}}}{R}}\sqrt {\left(\frac{dR}{d\tau}\right)^{2}+1},
\end{equation}
where $C$ is a constant. This may also be written in the form
\begin{equation}\label{eq7a}
\left(\frac{dR}{d\tau}\right)^2+1 -\frac{\left[4 \lambda J^2-\lambda
R^2\ln(R^2+J^2)-2 C R^2\right]^2}{4 R^2(R^2+J^2)} =0,
\end{equation}
and, therefore, the motion of the shell is identical to that of a
particle of unit mass in the potential,
\begin{equation}\label{eq8a}
 V(R)= \frac{1}{2}\left[1 -\frac{\left[4 \lambda J^2-\lambda
R^2\ln(R^2+J^2)-2 C R^2\right]^2}{4 R^2(R^2+J^2)}\right]
\end{equation}
with vanishing total energy. Notice that $C$ is not this energy and,
therefore, the form of $V(R)$ will be different for different
solutions. Nevertheless, (\ref{eq7a}) implies that if there are
suitable choices of the parameters for which the potential $V(r)$
has a negative minimum, the shell may execute a periodic motion. Let
us first find the conditions under which this may happen. We look
for equilibrium points (static solutions) where $\dot{R}=0$, and
$\ddot{R}=0$. Let $R=R_0$ be that point, then, from (\ref{eq6a}), we
have,
\begin{equation}\label{eq9a}
    \lambda = \frac{J^2 R_0 \sqrt{R_0^2+J^2}}{(R_0^2+2J^2)^2}
\end{equation}
which may also be considered as an equation for $R_0$, given
$\lambda$ and $J$,

To check for stable equilibrium points we set
$R(\tau)=R_0+\xi(\tau)$, replace in (\ref{eq6a}), and expand to
first order in $\xi(\tau)$. We find,
\begin{equation}\label{eq11a}
\frac{d^2\xi}{d \tau^2} = -\frac{[2 R_0^4- (R_0^2+2
J^2)J^2]J^2}{R_0^2(R_0^2+J^2)^2 (R_0^2+2 J^2)} \xi
\end{equation}

Then, the static solution will be stable for $R_0^2/J^2 >
(1+\sqrt{17})/4$, ($R_0/J > 1.1317...$) and unstable otherwise (see also \cite{hamity}).

The somewhat complex form and dependence on its parameters of $V(R)$
makes a general analysis of the possible motions based on
(\ref{eq7a}) rather difficult. We notice, however, that we have,
\begin{equation}
\label{eq12aa}
  V(R) = -(\lambda \ln(R)+C)^2/2+ {\cal{O}}(R^{0}) \; \;\; , \;\;\; R \to \infty
  \end{equation}
and
\begin{equation}
\label{eq13aa}
  V(R) = -2 \lambda^2 R^{-2}+ {\cal{O}}(R^{0}) \; \;\; , \;\;\; R \to 0
\end{equation}
and therefore we have unbounded motions for sufficiently large $R$
and collapsing motions for sufficiently small $R$. Moreover, the
equation,
\begin{equation}\label{eq14aa}
  4 \lambda J^2-\lambda R^2\ln(R^2+J^2)-2 C R^2 = 0
\end{equation}
has a real root with $R = R_m >0$ for any real $J$ and $C$, and
$\lambda>0$. But for $R=R_m$ we have $V(R) =1/2$, which is also the
maximum possible value of $V(R)$, and, therefore, the collapsing and
unbounded motion regions are separated at least by a "forbidden"
gap. Depending on the values of the parameters, $V(r)$ may contain
two "forbidden" gaps, where $V(r) > 0$, and periodic motions are
possible in the region between these gaps. Figure 1 provides some
explicit examples of these cases. They will be explored in more
detail in the next Sections.

\begin{figure}
\centerline{\includegraphics[height=12cm,angle=-90]{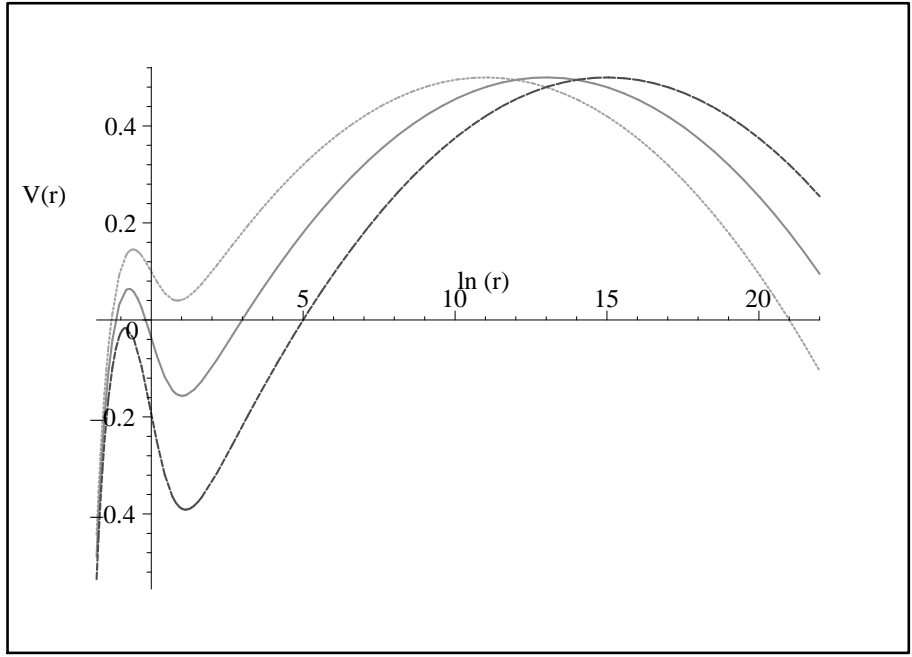}}
\caption{Plots of $V(r)$ as a function of $\ln(r)$. We have taken
$J=1$, and $\lambda=0.1$ in all cases. The solid curve, for
$C=-1.3$, contains a region where periodic motions are possible.
This region is absent for both the dotted ($C=-1.1$), and the dashed
($C=-1.5$) curves. In all cases we have a collapsing region, for
sufficiently small $r$, and unbounded region, for sufficiently large
$r$.}
\end{figure}

To close this Subsection we remark also that the evolution equation
(\ref{eq6a}) has the following scaling property: if we introduce the
function $\tilde{R}(\eta)$, such that,
\begin{equation}\label{eq12a}
\tilde{R}(\eta) \equiv \frac{1}{J} R(J \eta)
\end{equation}
we have
\begin{equation}\label{eq13a}
\frac {d^{2}\tilde{R}}{d{\eta}^{2}} +{\frac { \sqrt { \left( {\frac
{d \tilde{R}}{d\eta}} \right) ^{2}+1} \left(  \tilde{R}^{2}+2
\right) ^{2}\lambda}{ \left( \tilde{R}^{2}+1 \right) ^{3/2}
\tilde{R}^{2}}}-{\frac { \left( {\frac {d \tilde{R}}{d\eta}}
\right) ^{2}+1}{\tilde{R}
 \left(  \tilde{R}^{2}+1 \right) }} = 0
\end{equation}
and, therefore, all the types of motions, up to scalings, are
determined by the (adimensional) parameter $\lambda$.

\subsection{Compatibility with the field equations}

So far we have only considered Eq. (\ref{eq6a}). The full set of
field equations includes also $\psi$, $\gamma$ and the junction
conditions, and there is, a priori, no guarantee that the only
solutions of (\ref{eq6a}) compatible with these are the static ones.
In particular, the condition $\psi_{\Sigma}=0$ implies,
\begin{equation}\label{eq14a}
\psi^{+}_{,r}(R(\tau),t^+(\tau))\dot{R}
+\psi^{+}_{,t}(R(\tau),t^+(\tau)) X^+ = 0
\end{equation}
which, together with (\ref{eq8}), and (\ref{eq5a}), determine
$\psi^{+}_{,r}(R(\tau),t^+(\tau))$ and
$\psi^{+}_{,t}(R(\tau),t^+(\tau))$ in terms of $R(\tau)$. After some
simplifications, and using (\ref{eq7a}), we find,
\begin{eqnarray}
\label{eq15a}
\psi^+_{,r}(R(\tau),t^+(\tau)) &=& -\frac{\lambda [4\lambda J^2 +
8 \lambda R^2 +\lambda R^2 \ln(R^2+J^2) +2 C R^2]}{R (R^2+J^2)[1+4\lambda^2 \ln(R^2+J^2)+
8 \lambda C +16 \lambda^2]} \nonumber \\
\psi^+_{,t}(R(\tau),t^+(\tau))  &=& \frac{2 \lambda
\dot{R}}{\sqrt{R^2+J^2}[1+4\lambda^2 \ln(R^2+J^2) +8 \lambda C +16
\lambda^2]}
\end{eqnarray}
Similarly, from (\ref{eq5a}), we have,
\begin{equation}\label{eq16a}
\gamma^+(R(\tau),t^+(\tau)) = -\frac{1}{2} \ln [1+4\lambda^2
\ln(R^2+J^2)+8 \lambda C +16 \lambda^2]
\end{equation}
From this equation we may compute,
\begin{equation}\label{eq17a}
\frac{d \gamma^+}{d\tau} = \gamma^+_{,r} \dot{R}+ \gamma^+_{,t} X^+
\end{equation}
and we can check that if we replace (\ref{eq3}) on the right, and
then use (\ref{eq15a}), we get the same expression as that obtained
by computing the left hand side of (\ref{eq17a}) using
(\ref{eq16a}). We conclude that the restriction to a flat interior
is compatible with the dynamics of $\gamma$ on the shell, even in the non
stationary case.

Now we could compute in principle $\psi$  (and then $\gamma$) outside the shell. We notice
that, provided $R(\tau)$ satisfies some suitable conditions, to be
considered below, if we only imposed $\psi_{\Sigma}=0$, since
$R(\tau)$ is given, then we would get for $\psi$ a wave equation
with a well defined boundary condition  . But in this case both
$\psi_{,r}$ and $\psi_{,t}$ are given in the boundary, and it is not
clear that in this case we may get any non trivial solution. To
analyze this problem we notice that  in (\ref{eq2}) we may consider
$r$ as the ``time'' variable, and $t$ as the ``space'' variable
\cite{reula}, as shown in Figure 2. Then, the problem can be posed
as that of finding the evolution (in $r$) of $\psi$, for initial
data ($\psi_{\Sigma}=0, \psi_{,r}$) on the (one dimensional) surface
${\cal{S}}$,
\begin{figure}
\centerline{\includegraphics[height=12cm,angle=-90]{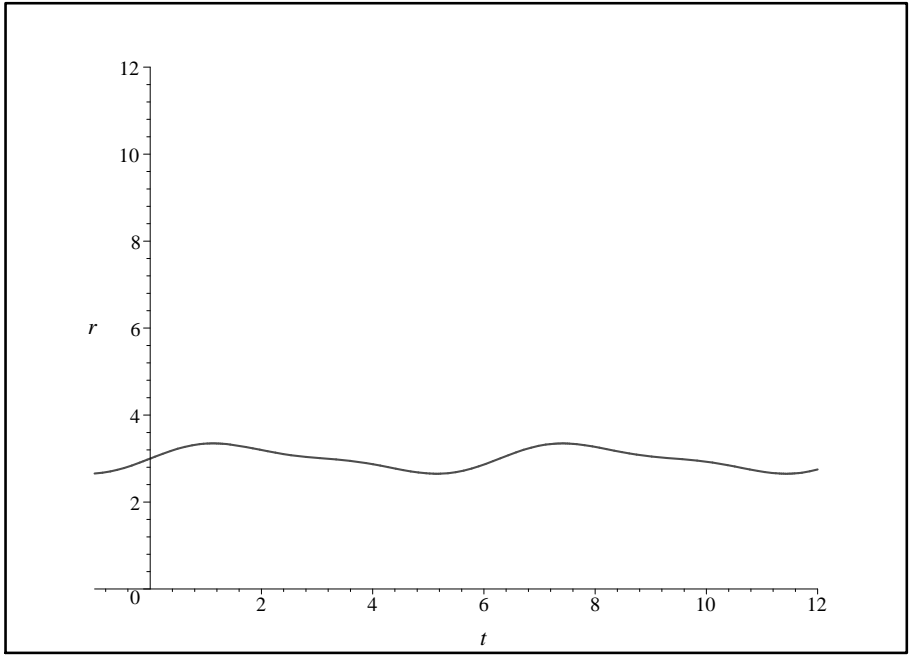}}
\caption{Evolving $\psi_+(t,r)$ from ``initial data"
$(\psi=0,\psi_{,r})$ on the curve $(r=R(\tau),t=t_+(\tau))$,
represented by the thick curve in the Figure. $\psi_+(t,r)$ is
evolved into the region above this curve. }
\end{figure}
\begin{equation}\label{eq18a}
 t = \eta \;\;\; ,\;\;\; r = R(\tau(\eta))
\end{equation}
where $\eta$ is a parameter, and $\tau(t)$ is obtained by inverting
(\ref{eq5a}). The problem is well posed provided that ${\cal{S}}$ is
a Cauchy surface, and this requires that the tangent vector to
${\cal{S}}$ be ``space like'', that is, $(dt/d\eta)^2
-(dr/d\eta)^2>0$, or
\begin{equation}\label{eq19a}
 \left(X^+\right)^2 > (\dot{R})^2
\end{equation}
But, from (\ref{eq7}), this is always satisfied. We might,
therefore, conclude that the field equations have solutions for any
$R(\tau)$ that is a solution of (\ref{eq6a}). However, we must also
require that the data $\psi_{,r}(R(\tau))$ be non singular, but, as
can be seen from (\ref{eq15a}), this may not always be the case,
because the factor $[1+4\lambda^2 \ln(R^2+J^2)+8 \lambda C +16
\lambda^2]$ in the denominator in (\ref{eq15a}) may vanish for some
finite $R \neq 0$. We notice here that solving (\ref{eq14aa}) for
$\ln(R_m^2+J^2)$ we get,
\begin{equation}\label{eq19aa}
1+4\lambda^2 \ln(R_m^2+J^2)+8 \lambda C +16 \lambda^2 = 1+16\,{\frac
{ \left( {R_{{m}}}^{2}+{J}^{2} \right) {\lambda}^{2}}{{R_
{{m}}}^{2}}}
\end{equation}
Since the left hand side is a monotonically increasing function of
$R$, this implies that the denominator is always positive for $R >
R_m$, and, therefore, this problem does not arise for the unbounded
solutions of the previous Subsection. Periodic motions are only
possible if the potential has, besides that for $R = R_m$ with
$V(R_m)=1/2$, another maximum for say $R=R_p$, with $V(R_p) >0$. For
this maximum we would have,
\begin{equation}\label{eq20a}
\left.\frac{dV}{dR}\right|_{R=R_p} =- \left.\frac{F_1 F_2}{4 R^3
(R^2+J^2)^2}\right|_{R=R_p} =0
\end{equation}
where,
\begin{equation}\label{eq21a}
F_1 = \lambda R^2\ln(R^2+J^2) +2CR^2-4 \lambda J^2
\end{equation}
and,
\begin{equation}\label{eq22a}
F_2 = {R}^{2}\lambda\,\ln  \left( {R}^{2}+{J}^{2} \right) {J}^{2}+2\,\lambda
\, \left( 2\,{J}^{4}+{R}^{4} \right) +2\,{R}^{2}{J}^{2} \left( C+4\,
\lambda \right)
\end{equation}

The first factor in (\ref{eq20a}), $F_1$, vanishes for $R=R_m$. We
can check that this is its only zero for $R > 0$ by noticing that,
\begin{equation}\label{eq23a}
\left.\frac{dF_1}{dR}\right|_{R=R_m} =2\,{\frac {\lambda\, \left(
{R_m}^{2}+2\,{J}^{2} \right) ^{2}}{R_m \left( {R_m}^{2}+{J}^{2}
\right) }} > 0
\end{equation}
and, therefore, the equation $F_1=0$ can have only one root for
$R>0$. Then, any extremum of $V(R)$ other than that for $R=R_m$ must
come from the vanishing of the second factor, $F_2$. It can be shown
that $F_2$ can have zero, one or two roots depending on the
parameters and that each root $R_p$ must satisfy $R_p<R_m$.  If
$F_2$ has no root or only one, $V(r)$ has only one maximum (at
$R_m$, the root of $F_2$ being a saddle point) and there are no
periodic motions. If $F_2$ has two roots, $V(r)$ has two maxima and
a minimum between them (a maximum and a minimum at the roots of
$F_2$ and another maximum at $R_m$). In this case we can have
periodic motions only if $V>0$ at the first maximum and $V<0$ at the
minimum, and this depends non trivially on the choice of parameters.
We can, nevertheless, obtain a useful result as follows. At a root
$R_p$ of $F_2$ we have,
\begin{equation}\label{eq24aa}
\ln  \left( {R_{{p}}}^{2}+{J}^{2} \right) =-2{\frac {2\,{J}^{4}+{R_{
{p}}}^{4}+4{R_{{p}}}^{2}{J}^{2}}{{R_{{p}}}^{2}{J}^{2}}}-2{\frac {C
}{\lambda}}
\end{equation}
and, therefore, we may write,
\begin{equation}\label{eq24a}
V(R_p) = \frac{1}{2}\left( 1 -\frac{ \lambda^2(R_p^2+2 J^2)^4}{J^4
R_p^2(R_p^2+J^2)}\right)
\end{equation}
Then, if $V(R_p)$ is a positive maximum we must have,
\begin{equation}\label{eq25a}
{\lambda}^{2}={\frac {\eta\,{J}^{4}{R_p}^{2} \left( {R_p}^{2}+{J}^{2}
 \right) }{ \left( {R_p}^{2}+2\,{J}^{2} \right) ^{4}}}
\end{equation}
with $0 \leq \eta < 1$. Replacing in the problematic factor in the
denominator of (\ref{eq15a}) we have,
\begin{equation}\label{eq26a}
 1+4{\lambda}^{2}\ln  \left( {R_p}^{2}+{J}^{2} \right) +8C\lambda+16
{\lambda}^{2}=1-8{\frac {{J}^{2} \left( {R_p}^{2}+{J}^{2} \right)
 \left( {R_p}^{4}+2{R_p}^{2}{J}^{2}+2{J}^{4} \right) \eta}{ \left( {R_p}
^{2}+2{J}^{2} \right) ^{4}}}
\end{equation}
The right hand side of (\ref{eq26a}) is a linear function of $\eta$.
It is equal to $1$ for $\eta=0$ and to $ {R_p}^8/({R_p}^2+{J}^2)^4$
for $\eta=1$. Since the left hand side of (\ref{eq26a}) is a
monotonic function of $R$, it is positive at any positive maximum of
$V(R)$ and the oscillating solutions always take place at $R>R_p$
(after the first positive maximum), we conclude that (\ref{eq15a})
is well defined and finite for any periodic motion.

As indicated, there are also collapsing solutions with a flat
interior. In this case we find that
${\psi^+}_{,r}(R(\tau),t^+(\tau))$ is singular either at some finite
$R$ or at $R=0$. It can be checked that these singularities occur
for finite $\tau$ and $t^{\pm}$. Since the solutions are symmetric
in $\tau$, this implies that the evolution has singularities both at
some finite time in the past and in the future, and, therefore, by
causality they extend only to some bounded region in $r$. We do not
analyze further these solutions as they do not seem to be physically
interesting.

In the following Sections we consider linearized solutions
corresponding to infinitesimally small departures from the static
stable solutions, both for flat and for empty regular interiors.

\section{Linearized periodic solutions with a flat interior}

Let us assume that $R= R_0$ corresponds, for some suitable $J$, to a
stable static solution with $\lambda$ given by (\ref{eq9a}). For
this solution we have $\psi_-=0$, $\psi_{\Sigma}=0$, and
$\gamma_-=0$, with $\psi_+$ and $\gamma_+$ given by
(\ref{eq1a},\ref{eq3a}). We consider now a perturbation of the
static solution such that the interior remains flat. This means that
we keep $\psi_-=0$, $\psi_{\Sigma}=0$, and $\gamma_-=0$, but for the
other dynamic variables we introduce now a time dependence by
setting,
\begin{eqnarray}
\label{eq1b}
R(\tau) &=& R_0 +\xi(\tau) \nonumber \\
\psi_+(r,t_+) &=& - \kappa \ln(r/R_0)+\bar{\psi}(r,t_+) \\
\gamma_+(r,t_+) &=& \gamma_0 +\kappa^2 \ln(r/R_0) +\bar{\gamma}(r,t_+) \nonumber
\end{eqnarray}
with $\kappa$ and $\gamma_0$ given by (\ref{eq3a}) with $R=R_0$, and
consider the linearized field equations that result from expanding
to first order in $\xi$, $\bar{\psi}$ and $\bar{\gamma}$. To this
order $\xi$ satisfies (\ref{eq11a}). If we define,
\begin{equation}\label{eq2b}
\Omega^2_0 = \frac{[2 R_0^4- (R_0^2+2 J^2)J^2]J^2}{R_0^2(R_0^2+J^2)^2 (R_0^2+2 J^2)}
\end{equation}
the solution of (\ref{eq11a}) can be written as,
\begin{equation}\label{eq3b}
\xi(\tau) = \xi_0 e^{i \Omega_0 \tau}
\end{equation}
In accordance with (\ref{eq7}), and (\ref{eq5a}), and expanding to
first order in $\xi_0$, we have,
\begin{equation}\label{eq4b}
\frac{dt_+}{d\tau}  = \frac {R_0^4}{(R_0^2+2 J^2)^2} +\frac{ 4 J^4 \xi_0 e^{i \Omega_0 \tau}}
{(R_0^2+2J^2)^2R_0}
\end{equation}
Actually, the last term on the right in (\ref{eq4b}) contributes in
all relevant equations only to second order, and, therefore, we may
set,
\begin{equation}\label{eq5b}
t_+ = \frac {R_0^4}{(R_0^2+2 J^2)^2} \tau
\end{equation}
when appropriate. Similarly, we may set $t_- = \tau$. We also
define, for convenience,
\begin{equation}\label{eq6b}
\Omega_2 = \frac  {(R_0^2+2 J^2)^2}{R_0^4} \Omega_0
\end{equation}
We look now for solutions of $\psi$ and $\gamma$ with the same
periodicity as $R(\tau)$. On account of (\ref{eq2}) and (\ref{eq5b})
the general solution for $\psi$ will be then of the form,
\begin{equation}\label{eq7b}
\psi_+(t_+,r)=-\frac{2J^2}{R_0^2}
\ln\left(\frac{r}{R_0}\right)+\left(A_2 J_0(\Omega_2 r)+B_2
Y_0(\Omega_2 r)\right)  e^{i \Omega_2 t_+}
\end{equation}
where $A_2$ and $B_2$ are constants, and $J_0$, and $Y_0$ are Bessel
functions. Then the junction conditions on the shell, and the
condition $\psi_{\Sigma}=0$ are satisfied (to first order in
$\xi_0$) if,
\begin{eqnarray}
\label{eq8b}
  A_2 &=& -\pi J^2
  \left[\frac{J^2 (4 J^2R_0^2+4 J^4-R_0^4)}{(R_0^2+J^2)R_0^7} Y_0(\Omega_2 R_0)+
\frac{ \Omega_0 (R_0^2+2J^2)^2}{R_0^6} Y_1(\Omega_2 R_0)\right]\xi_0 \nonumber \\
  B_2 &=& \pi J^2\left[
  \frac{J^2 (4 J^2R_0^2+4 J^4-R_0^4)}{(R_0^2+J^2)R_0^7} J_0(\Omega_2 R_0)+
   \frac{\Omega_0 (R_0^2+2J^2)^2}{R_0^6} J_1(\Omega_2 R_0)\right]\xi_0
\end{eqnarray}
Similarly, again to first order in $\xi_0$, we find
\begin{equation}\label{eq9b}
\gamma_+(t_+,r)= \gamma_0 + \frac{4J^4}{R_0^4}
\ln\left(\frac{r}{R_0}\right) - \frac{4J^2}{R_0^2}\left(A_2
J_0(\Omega_2 r)+B_2 Y_0(\Omega_2 r)\right)  e^{i \Omega_2 t_+}
\end{equation}
where,
\begin{equation}\label{eq10b}
\gamma_0 =\ln\left[ \frac  {(R_0^2+2 J^2)^2}{R_0^4} \right]
\end{equation}

Summarizing, we see that given appropiate values of $R_0$ and $J$,
we can find a complete solution, at the linearized level, where both
the motion of the shell and the radiative modes of the fields are
periodic in their respective times. For this type of solutions the
period is a definite function of $R_0$ and $J$, in correspondence
with the idea of a ``perturbation" of a stable equilibrium static
configuration, characterized by $R_0$ and $J$, with the departure
from equilibrium being given by the arbitrarily small parameter
$\xi_0$. Finally we remark that in the limit $J^2 <<R_0^2$ we have,
\begin{equation}\label{eq11b}
\Omega_0^2 \simeq \frac  {2 J^2}{R_0^4}
\end{equation}
that is, $\Omega_0$ approaches the value corresponding to small
oscillations of the shell in the Newtonian limit. At first sight it
would appear that this should be the {\em natural} frequency of
oscillation of the shell, and that the effect of the coupling to the
gravitational radiation modes should introduce only a small
departure, such as damping, from the Newtonian case. However, as we
shall show in the next section, this is only a special case
resulting from the assumption of a flat interior, and the behaviour
of the system is in general quite different from this expectation.

\section{Linearized periodic solutions with a regular interior}

We consider now the more general situation where the interior region
is empty but may contain gravitational radiation, imposing only the
condition of regularity on the symmetry axis $r=0$. We then set,
\begin{equation}\label{eq1d}
\psi_-(t_-,r)=  A_1 J_0(\Omega_- r)  e^{i \Omega_- t_-}
\end{equation}
Restricting again to linearized order we may set,
\begin{equation}\label{eq2d}
\gamma_-(t_-,r)= 0
\end{equation}
and, therefore, also to the appropriate order, we may also set,
\begin{equation}\label{eq3d}
 t_-(\tau)= \tau
\end{equation}
We assume again a perturbation around a stable equilibrium
configuration characterized by $R_0$ and $J$. We therefore take,
\begin{equation}\label{eq4d}
 R(\tau)= R_0+ \xi_0 e^{i \Omega \tau}
\end{equation}
and,
\begin{equation}\label{eq5d}
\psi_+(t_+,r)=-\frac{2J^2}{R_0^2}
\ln\left(\frac{r}{R_0}\right)+\left(A_2 J_0(\Omega_2 r)+B_2
Y_0(\Omega_2 r)\right)  e^{i \Omega_2 t_+}
\end{equation}
where $A_2$ and $B_2$ are constants, considered to be of first
order. To this order we then have,
\begin{equation}\label{eq6d}
\gamma_+(t_+,r)= \gamma_0 + \frac{4J^4}{R_0^4} \ln\left(\frac{r}{R_0}\right)
- \frac{4J^2}{R_0^2}\left(A_2 J_0(\Omega_2 r)+B_2 Y_0(\Omega_2 r)\right)
e^{i \Omega_2 t_+}
\end{equation}
where $\gamma_0$ is given by (\ref{eq10b}). A long calculation then
shows that consistency at first order of the equations requires
$\Omega_-=\Omega$,
\begin{equation}\label{eq7d}
\Omega_2 = \frac  {(R_0^2+2 J^2)^2}{R_0^4} \Omega =e^{\gamma_0} \Omega
\end{equation}
and,
\begin{equation}\label{eq8d}
t_+ = \frac {R_0^4}  {(R_0^2+2 J^2)^2}  \tau =e^{-\gamma_0} \tau
\end{equation}
Replacing now in (\ref{eq5}), (\ref{eq6}), and (\ref{eq9}), and
expanding to first order, we find a set of three linear independent
equations for $A_1$, $A_2$, $B_2$, and $\xi_0$. It turns out that a
convenient way of handling this system is to introduce a new
parameter $\alpha$ by the definition,
\begin{eqnarray}\label{eq9d}
 A_1 & = & \frac{R_0^2(R_0^2+2J^2)(R_0^2+J^2)^2\Omega^2
 +J^2(2J^4-2R_0^4+R_0^2J^2)}{\Omega^2R_0^2+1} \alpha
 \nonumber \\
 &=& \frac{R_0^2(R_0^2+2J^2)(R_0^2+J^2)^2(\Omega^2-\Omega_0^2)}{\Omega^2R_0^2+1} \alpha
\end{eqnarray}
where $\Omega_0^2$ is given by (\ref{eq2b}).
We then have,
\begin{equation}\label{eq10d}
\xi_0 = \frac{\left(R_0(R_0^2+2J^2)(R_0^2+J^2)\Omega J_1(\Omega
R_0)-J^2(2J^2+3R_0^2)J_0(\Omega R_0)\right)R_0^3}{\Omega^2 R_0^2+1}
\alpha
\end{equation}
and,

\begin{eqnarray}
\label{eq11d}
A_{{2}} & = & -\left[ \left[ \Omega \left( {R_0}^{2}+2{J}^{2} \right) ^{2} \left( {
R_0}^{2} \left( R_0^{2}+{J}^{2} \right) ^{2}{\Omega}^{2}-{J}
^{2} \left( {J}^{2}+2R_0^{2} \right)  \right) Y_{{1}} \left(
\Omega_{{2}}R_0 \right) \right.  \right. \nonumber \\ & &
\left.+2{J}^{2}Y_{{0}} \left( \Omega_{{2}}R_0
 \right) R_0 \left( R_0^{2}+{J}^{2} \right)  \left(
 \left( 4{J}^{4}+6R_0^{2}{J}^{2}+R_0^{4} \right) {
\Omega}^{2}-2{J}^{2} \right)  \right] J_{{0}} \left( \Omega R_0
 \right) \nonumber \\ & &
  - \left[ Y_{{0
}} \left( \Omega_{{2}}R_0 \right)  \left(  \left( R_0^{2}+2
{J}^{2} \right) ^{2}
\left(R_0^{2} \left( R_0^{2}+{J}^{2}
 \right) ^{2}{\Omega}^{2}
  - J^4 \right)-2J^2 R_0^6
 \right) \right. \nonumber \\ & &
\left. \left. -2{J}^{ 2}R_0\Omega \left( R_0^{2}+{J}^{2} \right)
\left( R_0^{2} +2{J}^{2} \right) ^{2}Y_{{1}} \left( \Omega_{{2}} R_0
\right)  \right]\Omega J_{{1}} \left( \Omega R_0 \right) \right]
\frac{\pi (R_0^2+2J^2) \alpha}{2 R_0^3(R_0^2\Omega^2+1)}
\end{eqnarray}

\begin{eqnarray}
\label{eq12d}
B_{{2}} & = &- \left[  \left( -2\,{J}^{2}Y_{{0}} \left( \Omega
_{{2}}R_{{0}} \right) R_{{0}} \left( {R_{{0}}}^{2}+{J}^{2} \right)
 \left(  \left( 4\,{J}^{4}+6\,{R_{{0}}}^{2}{J}^{2}+{R_{{0}}}^{4}
 \right) {\Omega}^{2}-2\,{J}^{2} \right) \right. \right.
 \nonumber \\ & &
 \left. -\Omega\, \left( {R_{{0}}}^{2
}+2\,{J}^{2} \right) ^{2} \left( {R_{{0}}}^{2} \left( {R_{{0}}}^{2}+{J
}^{2} \right) ^{2}{\Omega}^{2}-{J}^{2} \left( {J}^{2}+2\,{R_{{0}}}^{2}
 \right)  \right) Y_{{1}} \left( \Omega_{{2}}R_{{0}} \right)  \right)
J_{{0}} \left( \Omega\,R_{{0}} \right)
\nonumber \\ & &
+J_{{1}} \left( \Omega\,R_{{0}}
 \right) \Omega\, \left( Y_{{0}} \left( \Omega_{{2}}R_{{0}} \right)
 \left(   \left( {R_{{0}}}^{2}+2\,{J}^{2}
 \right) ^{2} \left( {R_{{0}}}^{2} \left( {R_{{0}}}^{2}+{J}^{2}
 \right) ^{2}{\Omega}^{2}-{J}^{4} \right)-2\,{J}^{2}{R_{{0}}}^{6}
 \right) \right.
\nonumber \\ & &
 \left. \left. -2\,{J}^{2}R_{{0}}
\Omega\, \left( {R_{{0}}}^{2}+{J}^{2} \right)  \left( {R_{{0}}}^{2}+2
\,{J}^{2} \right) ^{2}Y_{{1}} \left( \Omega_{{2}}R_{{0}} \right)
 \right)  \right]
\frac{ {\pi}\, \left( {R_{{0}}}^{2}+2\,{J}^{2} \right)
\alpha }
{2 \left( {R_{{0}}}^{2}{\Omega}^{2}+1 \right) {R_{{0}}}^{3}}
\end{eqnarray}

The main reason for displaying these, at first sight, not very
illuminating expressions for $A_1$, $\xi_0$, $A_2$ and $B_2$ is that
they explicitly show that given $R_0$ and $J$ corresponding to some
stable equilibrium configuration, i.e., to some real value for
$\Omega_0$, we have non trivial periodic solutions for the
linearized perturbations for {\em every} value of $\Omega$. Thus, we
reach the unexpected result that, at least perturbatively, we cannot
ascribe a particular period to motions close to the stationary
solution, as happens in the corresponding Newtonian dynamics. The
period of the motion can be arbitrary, depending entirely on the
field configuration. From a more physical point of view, this can be
interpreted by noticing that as the radius of the shell changes, the
change in the static part of the field (the $\ln(r)$ terms) is of
the same order of magnitude as the radiating part of the field that
this motion generates. Thus, as the shell moves away from its
stationary configuration, the motion is driven to essentially
similar extents by the static and the dynamic parts of the
gravitational field. In some sense then, the coupling of the shell
to the gravitational radiation modes is as strong as it can be, a
remarkable fact that shows that the dynamics of this system cannot be
approximated by a Newtonian dynamics plus post - Newtonian
corrections, as in the case of some more realistic models, where matter
is confined to a bounded region.

We have already analyzed the special case where the field inside the
shell vanishes, and found that this is possible only for a
particular value of $\Omega$, which, in the context of this more
general analysis, corresponds to the particular solution where
$A_1=0$. In fact it is straightforward to show that the solution for
$\Omega = \Omega_0$ reduces precisely to that of the previous
Section. But there is also, for instance, a particular set of solutions that display a
different type of unexpected behaviour. We may call these ``anti
resonances". They are described in the next section.

\section{Anti-resonances}

As the shell evolves in time, its physical radius is given by
$R_0(\tau) \exp(-\psi_{\Sigma}(\tau))$. For perturbations around an
equilibrium point, to linear order we then have,
\begin{equation}\label{eq1Ar}
R_0(\tau) e^{-\psi_{\Sigma}(\tau)} = R_0 + \left(\xi_0- A_1 R_0
J_0(\Omega R_0) \right)e^{i \Omega \tau}
\end{equation}

If we use now (\ref{eq9d}) and (\ref{eq10d}), we get,
\begin{eqnarray}
\label{eq2Ar}
 \xi_0- A_1 R_0 J_0(\Omega R_0)& = &
  \left[{R_0}^{3}\Omega\,J_{{1}} \left(\Omega\,R_0 \right)
 - \left( {J}^{2}+{R_0}^{2}{\Omega}^{2} \left( {R_0}^{2}+{J}^{2} \right)  \right) J_{{0}
} \left( \Omega\,R_0 \right) \right] \nonumber \\
 & & \times \frac {\,R_0 \left( {R_0}^{2}+2\,{J}^{2} \right)
 \left( {R_0}^{2}+{J}^{2} \right) \alpha
  }{{R_0}^{2}{\Omega}^{2}+1}
\end{eqnarray}
This implies that the physical radius of the shell remains constant
(to first order), and hence we have an {\em anti-resonance}, if
$\Omega$ is a solution of the equation,
\begin{equation}
\label{eq3Ar}
{R_0}^{3}\Omega\,J_{{1}} \left(\Omega\,R_0 \right)
 - \left( {J}^{2}+{R_0}^{2}{\Omega}^{2} \left( {R_0}^{2}+{J}^{2} \right)  \right) J_{{0}
} \left( \Omega\,R_0 \right) = 0
\end{equation}
It is easy to check that (\ref{eq3Ar}) has an infinite sequence of
solutions. Again it is remarkable that for these frequencies the
effects of the inner an outer radiation modes exactly compensate
each other and the shell remains motionless.

\section{The general behaviour of the periodic solutions}

The linearized solutions found in the previous sections have in
common the desirable feature that they contain a parameter that can
be made arbitrarily small, and thus they approach arbitrarily
closely the static solution. At least this is true for finite values
of $r$. In fact, looking at the form of (\ref{eq7b}) and
(\ref{eq9b}) we see that for large $r$ the solution appears to be
dominated by the static $\ln(r)$ terms, and, therefore, that the
solutions approach the static unperturbed background for large $r$.
A more accurate geometrical picture can be obtained by considering,
e. g., the Kretschmann invariant $K$, for large $r$. Using the forms
(\ref{eq7b}) and (\ref{eq9b}) we find,
\begin{eqnarray}\label{eq1f}
K & = & -\frac{16
\kappa(1+\kappa)(1+2\kappa)}{e^{\gamma_0}R_0^{-4\kappa(1+\kappa)}
r^{4+4\kappa(1+\kappa)}}\cos(\Omega_1 t_+)\sqrt{ \frac{2 \Omega_1^3
r^3}{\pi}}
\nonumber \\
& & \times \left[A_1\sin\left(\Omega_1 r
+\frac{\pi}{4}\right)-A_2\cos\left(\Omega_1 r
+\frac{\pi}{4}\right)\right] +{\cal{O}}\left(\frac{r^{1/2}}{
r^{4+4\kappa(1+\kappa)}}\right)
\end{eqnarray}
while for the background static metric we have,
\begin{equation}\label{eq2f}
K = \frac{16
\kappa^2(1+\kappa+\kappa^2)(1+\kappa)^2}{e^{4\gamma_0}R_0^{-4\kappa(1+\kappa)}
r^{4+4\kappa(1+\kappa)}}
\end{equation}
Thus, although in both cases $K \to 0$ for large $r$, in the
perturbed case $K$ is a factor of order $r^{3/2}$ larger than in the
static case, so this appears to indicate a larger and larger
departure between the perturbed and perturbed solutions as $r \to
\infty$. The consequences and meaning of this departure are not
clear. For instance, in the flat interior case, where we have the
same behaviour for $K$, we have shown that there are non
perturbative periodic solutions, and the linearized solutions should
approach those, and therefore, at least in that sense, the behaviour
(\ref{eq1f}) would be compatible with a perturbative treatment.
This, however, is not entirely correct. The reason is that if we
attempt to solve the field equations for $\psi_+$ and $\gamma_+$ to
second order in the periodic terms, $\gamma_+$ acquires terms of
order $r$ (rather than order $r^0$ as in the first order terms),
and, if these are included in $K$ the difference between the
solutions is now of order $r^3$.

There is nevertheless, another way to look at these solutions. As
indicated, they are indeed close to the static solution provided $r$
is not too large. We notice that the equations for  $\psi_+$ and
$\gamma_+$ are local and causal. In particular, $ \partial
\gamma_+/\partial t_+$ vanishes if $ \partial \psi_+/\partial t_+ =
0$. We may, therefore, consider the solution up to some large value
of $r$, say $r_b >>R_0$, cut off the periodic part for $r > r_b$,
and use this configuration as initial data for the system. On account of
causality, the shell will then  oscillate periodically for a time of
the order of $r_b$, while preserving the asymptotic structure of the
static solution, so that, in principle, we could have solutions that
are periodic for a time that is long as compared with the
oscillation period.

There is still another way to look at the linearized solutions that
is explored in the next section.

\section{The initial value problem}

An important problem related to the system under discussion is the
following. Suppose we have initial data that differs slightly from
that corresponding to the static solution. We then expect that the
evolution of that data will remain close to static solution and
that, therefore, a linearized treatment should be adequate. We
notice at this point that a linear superposition of linearized
periodic solutions will also be a linearized solution, although no
longer periodic. In fact, we may generalize this idea and write,
\begin{eqnarray}
\label{eq1g}
\psi_+(t_+,r) &=& -\kappa \ln(r/R_0) + \int_0^{\infty}\left[a_2(\Omega_+) J_0(\Omega_+ r)
+ b_2(\Omega_+) Y_0(\Omega_+ r)\right] e^{i \Omega_+ t_+} d \Omega_+   \nonumber \\
\gamma_+(t_+,r) &=& \gamma_0+\kappa^2 \ln(r/R_0)
-2 \kappa \int_0^{\infty}\left[a_2(\Omega_+) J_0(\Omega_+ r)
+ b_2(\Omega_+) Y_0(\Omega_+ r)\right] e^{i \Omega_+ t_+} d \Omega_+ \nonumber \\
\psi_-(t_-,r)
&=&  \int_0^{\infty}a_1(\Omega_-) J_0(\Omega_- r)  e^{i \Omega_- t_-} d \Omega_-  \\
\gamma_- &=& 0 \nonumber \\
R(\tau) &=& R_0 + \int_0^{\infty}\xi(\Omega) e^{i \Omega \tau} d
\Omega \nonumber
\end{eqnarray}
where the coefficients $a_1$, $a_2$, $b_2$, and $\xi$ are complex
functions of their arguments, and, as usual, it is understood that
we take the real part of the right hand side of (\ref{eq1g}). Since
each value of $\Omega$ is independent of the others, we may use the
results of the previous Section to solve for $a_2$, $b_2$, $\xi$ and
$a_1$ in terms of a different $\alpha$, for each $\Omega$, and it is
clear that we may choose $\alpha$ to be an arbitrary complex
function $\Omega$. In particular, considering the asymptotic
behaviour for large $r$ of the Bessel functions $J_0$ and $Y_0$, we
see that with an appropriate fall off for $\alpha(\Omega)$ as
$\Omega \to \infty$ we may control the corresponding fall off for
large $r$ of $\psi_+$ and $\gamma_+$, because the dependence on
$\Omega$ of both $a_2$ and $b_2$ is related by linearity to that of
$\alpha$. Since the expressions for $\psi_+$ and $\gamma_+$ have the
form of Fourier - Bessel transforms, in this case the radiative
parts should also fall off for large $|t|$, and these expressions
might represent a situation where for large negative $t$ the shell
is stationary, being subsequently perturbed by an incoming
gravitational radiation pulse, which eventually rebounds, leaving
the shell again in a stationary state.

Although the above reasoning is correct, it is not clear how we can
use it to solve the {\em initial value problem} for our system. To
begin with, assuming that, e.g., $\psi_+(0,r)$ is given, since the
range of $r$ is {\em not} $0 \leq r < \infty$, and we cannot impose
a priori boundary conditions for $r=R_0$, there appears to be no
well defined procedure for inverting (\ref{eq1g}) and computing,
say, $a_2$ and $b_2$. Nevertheless, since $\alpha$, and therefore,
$a_2$ and $b_2$, are complex, we actually have {\em two} arbitrary
real functions of $\Omega_+$ at our disposal for the construction of
$\psi_+(0,r)$, and, therefore, make the system satisfy arbitrary
initial data. We notice, however, that once $\alpha(\Omega)$ is
given, not only $a_2$ and $b_2$ are fixed, but also $a_1$, and
therefore, the data {\em inside} the shell, which should, from
causality, be independent of that {\em outside} the shell. The
answer to this conundrum is that the expression for  $\psi_+(0,r)$
on the right in (\ref{eq1g}) is overcomplete, because we only
require $\psi_+(0,r)$ in the range $R_0 \leq r < \infty$, and that
leaves the range $0 < r < R_0$ arbitrary. Since the expression in
(\ref{eq1g}) actually defines $\psi_+(0,r)$ also in the region $0 <
r < R_0$, there must be an infinite set of functions $a_2$ and $b_2$
that reproduce the data in $R_0 \leq r < \infty$, so in principle
there is room for arbitrary data $\psi_-(0,r)$ in $0 < r < R_0$.
Again, although this seems plausible, we do not have a proof of its
validity. The difficulty here is the lack of a self adjoint
formulation for the initial value formulation of the moving boundary
problem posed by the dynamics of our system. This problem will be
considered in detail elsewhere \cite{Ragle}.

To illustrate the points considered in this Section, we include as
an example, the case of an incoming pulse, its interaction with the
shell, and eventual rebound after this interaction. In this example
we set $R_0=4$, $J=1$, which implies $\lambda=0.0509...$, and
$\Omega_0= 0.0770...$. We also set,
\begin{equation}\label{eq2g}
{\alpha(\Omega)}=2\,{\frac {{{R_0}}^{3} \left(
{\Omega}^{2}{{R_0}}^{2}+1 \right) Q {{ e}^{-4\, \left( \Omega-2
\right) ^{2}}} }{ \left( {{ R_0}}^{2}+2 \,{J}^{2} \right) \pi }}
\end{equation}
with $Q=10^{-5}$. Replacing in (\ref{eq1g}) we obtain explicit
expressions for the dynamic variables of the problem, from which we
can view the evolution of the system. Details are given in Figs. 3,
4 and 5. In Figure 3 we have a plot of $R(\tau)-R_0$ as a function
of $\tau$, showing the incoming pulse region, for $\tau < 0$, and
the outgoing pulse region for $\tau >0$. The shell is essentially in
its equilibrium radius $R_0$ for either $\tau << -10$ and $ \tau >>
10$. Figure 4 is a plot of $\psi_-(t,r)$ in the region $0 \leq r\leq
R_0=4$, $ -10 \leq t \leq 10$. We notice the propagation of the
incoming pulse, an intermediate interference zone, formed by
incoming and outgoing waves, and the eventual fall off of the pulse
as it propagates outside. In Figure 5 we have a plot of
$\psi_+(t,r)$ in the region $R_0=4 \leq r\leq 15$, $ -10 \leq t \leq
10$. We can see the propagation of the incoming pulse towards the
shell, a zone near $t=0$ where most of the pulse has gone through
the shell, and its rebound and propagation away from the shell, for
$t >0$.

\begin{figure}
\centerline{\includegraphics[height=12cm,angle=-90]{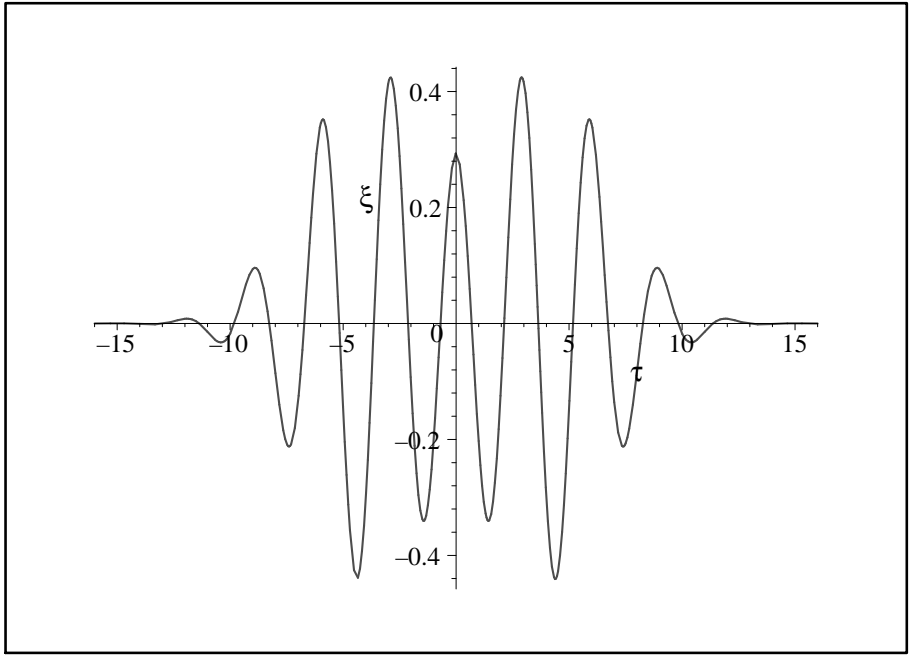}}
\caption{Plot of $\xi(\tau) =R(\tau)-R_0$ in the region $-10 \leq
\tau \leq 10 $. The region $\tau < 0$ is dominated by the incoming
pulse, while that for $\tau>0$ is dominated by the outgoing pulse,
resulting from the rebound of the pulse on the symmetry axis. The
shell is essentially in its equilibrium radius $R_0$ for either
$\tau << -15$ and $ \tau >> 15$.}
\end{figure}

\begin{figure}
\centerline{\includegraphics[height=12cm,angle=-90]{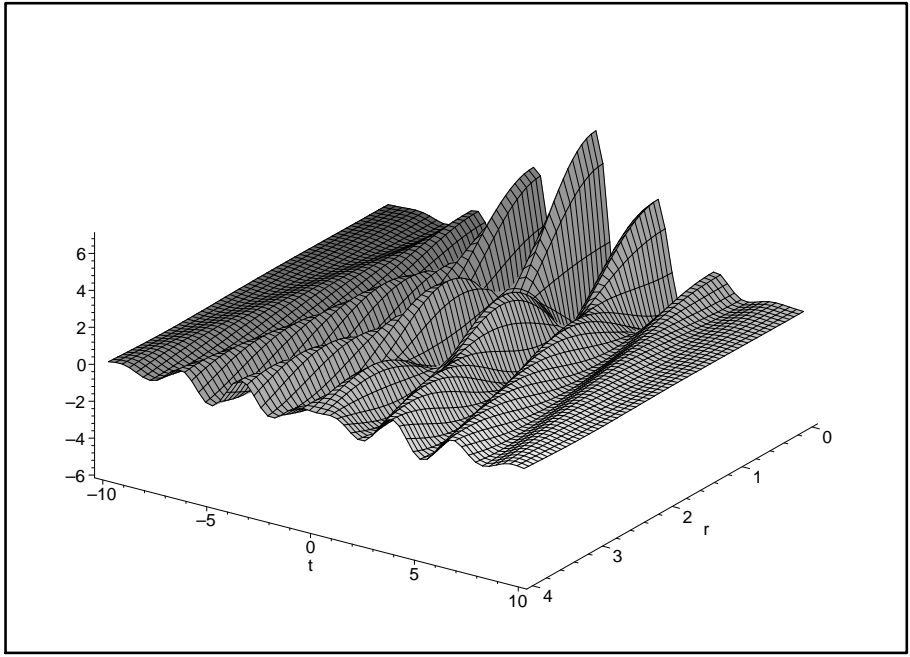}}
\caption{Plot of $\psi_-(t,r)$ in the region $0 \leq r\leq R_0=4$, $
-10 \leq t \leq 10$. We notice the propagation of the incoming
pulse, an intermediate interference zone, formed by incoming and
outgoing waves, and the eventual fall off of the pulse as it
propagates outside.}
\end{figure}

\begin{figure}
\centerline{\includegraphics[height=12cm,angle=-90]{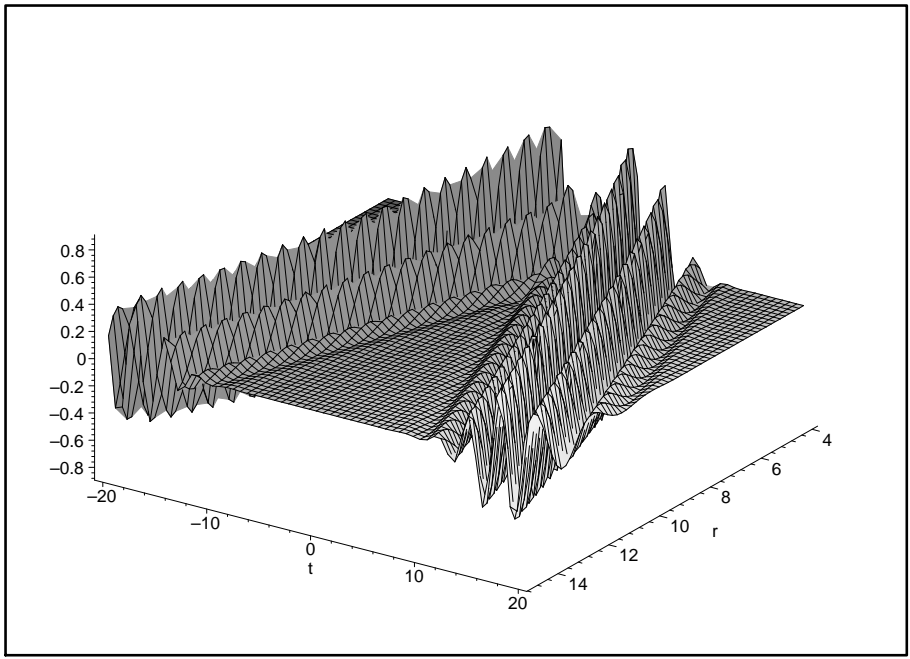}}
\caption{Plot of $\psi_+(t,r)$ in the region $4 \leq r\leq 15$, $
-20 \leq t \leq 20$. We can see the propagation of the incoming
pulse towards the shell, a zone near $t=0$ where most of the pulse
has gone through the shell, and the rebound and propagation of the
pulse away from the shell, for $t >0$.}
\end{figure}

In closing this Section we may ask what is the relation of this
construction, and its implied initial data, to the ``momentarily
static and radiation free" initial data of \cite{apostol}. (See
also \cite{nakao} for a different analysis of the meaning of this
type of data). To understand this we consider again the general
equations. Without loss of generality we may choose $\tau$, $t_-$,
and $t_+$ such that $\tau=0$ corresponds to $t_+ = t_-=0$, and then,
for the ``momentarily static and radiation free" initial data on
the surfaces $t_+ = t_-=\tau=0$, we would have,
\begin{eqnarray}
\label{eq3g}
  R(\tau) &=& R_1 + R_2\; \tau^2/2 + \dots \nonumber \\
  \psi_-(t_-,r) &=& P_1(r)\; t_-^2 + \dots \nonumber \\
  \gamma_-(t_-,r) &=& G_1(r)\; t_-^2 + \dots  \\
 \psi_+(t_+,r) &=& Q_1 \ln(r/R_1) + P_2(r)\; t_+^2 + \dots  \nonumber \\
  \gamma_+(t_+,r) &=& Q_1^2 \ln(r/R_1) +Q_2 + G_2(r)\; t_+^2 + \dots \nonumber
\end{eqnarray}
where $P_1$, $P_2$, $G_1$, and $G_2$ are some functions of $r$ that
are determined by the evolution, and whose explicit form is not
relevant, and dots indicate higher order (in $\tau$ and $t_{\pm}$)
terms. Replacing these expressions in the general equations of
motion we find that we must set,
\begin{eqnarray}
\label{eq4g}
  Q_{{1}} & = & {\frac {2 \lambda\,{R_{{1}}}^{2}}{4\,\lambda\, \left( {R_{{1}
}}^{2}+{J}^{2} \right) -R_{{1}}\sqrt {{R_{{1}}}^{2}+{J}^{2}}}} \nonumber \\
  Q_{{2}} & = & -\ln\left({\frac {R_{{1}}-4\,\lambda\,\sqrt {{R_{{1}}}^{2}+{J}^{2}
} }{R_{{1}}}}\right)  \\
 R_{{2}} & = & -{\frac { \left( 2\,{J}^{2}+{R_{{1}}}^{2} \right) ^{2}\lambda-
{J}^{2}\sqrt {{R_{{1}}}^{2}+{J}^{2}}R_{{1}}}{ \left( {R_{{1}}}^{2}+{J}
^{2} \right) ^{3/2}{R_{{1}}}^{2}}} \nonumber
\end{eqnarray}

We recall now that for a static solution we have,
\begin{equation}\label{eq5g}
\lambda={\frac {{J}^{2}\sqrt {{R_{{0}}}^{2}+{J}^{2}}R_{{0}}}{ \left( 2
\,{J}^{2}+{R_{{0}}}^{2} \right) ^{2}}}
\end{equation}
where $R_0$ is the equilibrium radius. For small departures from
this radius we may set,
\begin{equation}\label{eq6g}
R_1=R_0 + \xi
\end{equation}
where $\xi$ is constant. Replacing (\ref{eq5g}) and (\ref{eq6g}) in
(\ref{eq4g}), and expanding to first order in $\xi$ we find,
\begin{eqnarray}
\label{eq7g}
 Q_{{1}} & = & -2\,{\frac {{J}^{2}}{{R_{{0}}}^{2}}}+2\,{\frac {{J}^{4}
 \left( -{R_{{0}}}^{4}+4\,{J}^{2}{R_{{0}}}^{2}+4\,{J}^{4} \right) }{{R
_{{0}}}^{7} \left( {R_{{0}}}^{2}+{J}^{2} \right) }}\xi+O \left( {\xi}^
{2} \right)  \nonumber \\
 Q_{{2}}& = &-4\,\ln  \left( R_{{0}} \right) +2\,\ln  \left( 2\,{J}^{2}+{R
_{{0}}}^{2} \right) -4\,{\frac {{J}^{4}}{{R_{{0}}}^{5}}}\xi+O \left( {
\xi}^{2} \right)
  \\
R_{{2}} & = & {\frac {{J}^{2} \left( -2\,{R_{{0}}}^{4}+{J}^{2}{R_{{0}}}^{2}
+2\,{J}^{4} \right) }{ \left( 2\,{J}^{2}+{R_{{0}}}^{2} \right)
 \left( {R_{{0}}}^{2}+{J}^{2} \right) ^{2}{R_{{0}}}^{2}}}\xi+O \left(
{\xi}^{2} \right)
 \nonumber
\end{eqnarray}

Thus, for sufficiently small departures from the static stable
solution the ``momentarily static and radiation free" initial data
corresponds to adding a constant to $\psi_+(0,r)$, while leaving
$\psi_-(0,r)=0$. As indicated already, it appears possible, at least
in principle, to find coefficients in (\ref{eq1g}) that correspond
to this initial data, and that could be used to study its evolution,
but this problem has not yet been solved. As a final comment, we
notice that this does not correspond to perturbations of essentially
compact support, and therefore, one would expect that the
corresponding integrals would contain some singular expression. This
may not be a problem, because to analyze the motion for finite times
near this initial data we may simply cut off the perturbation for
large $r$, which, from causality, would not modify the solution for
times of the order of $r$. We, nevertheless, refer to \cite{nakao}
for a different analysis of this problem.

\section{Comments}

In this paper we have presented an analysis of the dynamics of a
self gravitating cylindrical thin shell of counter rotating dust
particles. This analysis provides several new and to a certain
extent unexpected results. In particular we show that there exists a
family of solutions where the interior of the shell remains flat at
all times. For this family the equation of motion for the radius of
the shell decouples from the radiative modes. We find a first
integral for this equation and show it to be equivalent to that of a
particle in a one dimensional time independent potential for a
certain value of its total energy. Depending on the constants of the
motion we have collapsing, periodic or unbounded solutions. We
further analyze under what conditions these solutions are consistent
with the field equations for the gravitational modes, and show that
there are consistent periodic solutions for the full system. We
consider next the dynamics of the system close to a stable static
solution in the linearized approximation, where we assume that the
interior is regular. The first unexpected result is that we have non
trivial solutions for any possible frequency, and that all these
modes are stable. The only role played by the Newtonian frequency
(corresponding to Newtonian dynamics of the shell) is that it is the
only frequency for which the interior is flat. We thus reach the
conclusion that the system has no ``natural" oscillating frequency
that would be slightly modified by the coupling to the radiative
modes, but, rather, we have a system where this coupling is ``as
strong as it can be", fully determining the behaviour of the shell.
We find also an infinite family of ``anti-resonances", where the
physical radius of the shell is constant (to first order).

The fact that we have an infinite set of modes suggests that these
modes could be used to solve the initial value problem (in the
linearized approximation). In fact, we can formally write an
arbitrary solution of the field equations as an integral transform
involving Bessel functions. Unfortunately, we have not found a way
to invert this transforms in such a way that they can be written in
terms of arbitrary initial data, although this seems to be possible in
principle. We discussed the reasons that make this problem special,
and provided a particular example to illustrate some features of the
general solution.

\section*{Acknowledgments}

This work was supported in part by grants from CONICET (Argentina).
MAR is supported by CONICET. RJG acknowledges partial support from
CONICET


\begin{thebibliography}{99}
\bibitem{note1} It was assumed in \cite{apostol} that the MSRF data
would always evolve towards a static equilibrium configuration,
something that it is not always true, as shown in \cite{nakao}
\end{thebibliography}
\end{document}